# Gain dynamics of a free-space nitrogen laser pumped by circularly polarized femtosecond laser pulses

Jinping Yao[1], Hongqiang Xie[1,3], Bin Zeng[1], Wei Chu[1], Guihua Li[1,3], Jielei Ni[1], Haisu Zhang[1,3], Chenrui Jing[1,3], Chaojin Zhang[1], Huailiang Xu[2,4], Ya Cheng[1,5], and Zhizhan Xu[1,6]

[1] *State Key Laboratory of High Field Laser Physics, Shanghai Institute of Optics and Fine Mechanics, Chinese Academy of Sciences, Shanghai 201800, China*
[2] *State Key Laboratory on Integrated Optoelectronics, College of Electronic Science and Engineering, Jilin University, Changchun 130012, China*
[3] *University of Chinese Academy of Sciences, Beijing 100049, China*
[4] *huailiang@jlu.edu.cn*
[5] *ya.cheng@siom.ac.cn*
[6] *zzxu@mail.shcnc.ac.cn*

**Abstract**: We experimentally demonstrate ultrafast dynamic of generation of a strong 337-nm nitrogen laser by injecting an external seed pulse into a femtosecond laser filament pumped by a circularly polarized laser pulse. In the pump-probe scheme, it is revealed that the population inversion between the $C^3\Pi_u$ and $B^3\Pi_g$ states of $N_2$ for the free-space 337-nm laser is firstly built up on the timescale of several picoseconds, followed by a relatively slow decay on the timescale of tens of picoseconds, depending on the nitrogen gas pressure. By measuring the intensities of 337-nm signal from nitrogen gas mixed with different concentrations of oxygen gas, it is also found that oxygen molecules have a significant quenching effect on the nitrogen laser signal. Our experimental observations agree with the picture of electron-impact excitation.

## 1. Introduction

In the past decades, femtosecond laser filamentation has been a hot research subject in strong field laser physics and ultrafast nonlinear optics because it is not only involved in rich physical phenomena such as high harmonic generation [1], terahertz radiation [2], molecular orientation and alignment [3], alignment-dependent fluorescence emission [4,5], but also triggers a series of promising applications including spectral broadening and pulse compression [6], lightning control [7], remote sensing [8,9], etc. In particular, a so-called air laser induced by femtosecond laser filamentation has recently attracted significant attention, which can operate in either forward or backward directions. The ultrafast laser induced air laser can be realized by either amplified spontaneous emission (ASE) in molecular nitrogen ($N_2$) [10,11] or seed amplification in nitrogen molecular ions ($N_2^+$) [12–20]. Up to the present, the mechanisms behind both the above-mentioned two types of nitrogen-based air lasers have not been completely clarified.

Recently, it was found that the strong neutral-nitrogen-based free-space lasers at 337 nm can be generated in backward direction by an 800-nm circularly polarized femtosecond laser [21] or in forward direction by a 1053-nm linearly polarized picosecond laser [22]. These findings suggest that inelastic collision between the electrons produced by the pump pulses and neutral nitrogen molecules plays a key role in producing the population inversion [21,22]. In particular, the backward laser achieved with the circularly polarized femtosecond laser pulses facilitates further investigation on the gain dynamics of the free-space 337-nm laser generated in neutral nitrogen molecules by using a pump-probe method, as the pump laser source used is a typical 800-nm Ti:sapphire laser with a pulse energy of ~10 mJ and the pulse duration of ~50 fs [21]. The short duration of the pump pulse allows us to perform the pump-probe investigation on the gain dynamics with a high temporal resolution.

Previously, we have developed a pump-probe technique for investigating the lifetime of population inversion in air lasers using $N_2^+$ as the gain medium [13–15,23]. Thanks to the assistance of an external seed pulse (i.e., probe pulse), the signal of the air laser can be dramatically enhanced [13–15]. In addition, the evolution of rotational wave packets of $N_2^+$ can be observed with a high temporal resolution [14,23]. We expect that most of these advantages will remain when the pump-probe scheme is employed for producing the free-space $N_2$ laser.

In this work, to shed more light on the underlying mechanism behind the ultrafast population inversion in the neutral nitrogen molecules, we generate the free-space 337-nm nitrogen laser using the pump-probe scheme. Our measurements reveal the built-up time and the decay time of the population inversion between the $C^3\Pi_u$ and $B^3\Pi_g$ states of $N_2$ in the generation of the 337-nm laser at different nitrogen gas pressures. The experiment also shows that mixing of nitrogen gas with oxygen gas can significantly alter the gain dynamics of the 337-nm laser.

## 2. Experimental setup

The schematic diagram of the experimental setup is shown in Fig. 1. The femtosecond laser pulses (~40 fs, 1 kHz, 800 nm, linearly polarized) from a commercial Ti:sapphire laser (Legend Elite-Duo, Coherent, Inc.) were split into two beams. One beam with a pulse energy of ~4 mJ was used as the pump to generate a plasma filament in the nitrogen gas and to build up the population inversion between the $C^3\Pi_u$ and $B^3\Pi_g$ states of $N_2$. A quarter-wave plate (QWP) was placed in the pump beam to change the polarization of the pump pulse from linear to circular. The other beam with a pulse energy of ~2.8 mJ was first directed into a 20-mm-long BK7 glass to produce the supercontinuum light on the blue side of the spectrum. To generate the strong supercontinuum emission in the BK7 glass, the beam diameter of the probe was reduced to ~4.5 mm by a telescope system. The supercontinuum light was then frequency-doubled in a 2-mm-thickness β-BaB$_2$O$_4$ (BBO) crystal. In this way, the spectrum of the second harmonic can be continuously tuned by adjusting the phase-matched angle of the BBO crystal. A dichroic mirror (DM) with a high reflectivity around 800 nm and a high transmission in the 290–660 nm spectral range (Filter 1) was used to select the frequency-doubled light at 337 nm, which was then employed as the probe. The pump and probe pulses were combined using another DM with the same parameters, and then collinearly focused by an *f*=30cm lens into a gas chamber filled with nitrogen gas at different pressures. The second DM was arranged at nearly zero degree in order to avoid the reduction of circular-polarization degree. The time delay between the pump and the probe was controlled by a motorized linear translation stage, which was inserted into the probe beam. All the light beams after the gas chamber were collimated by an *f*=40cm lens. After passing through an interference filter centered at ~340 nm with a bandwidth of 10 nm (Filter 2) and an attenuator, the forward spectra of the nitrogen laser were recorded by a grating spectrometer (Shamrock 303i, Andor) equipped with a 1200 grooves/mm grating.

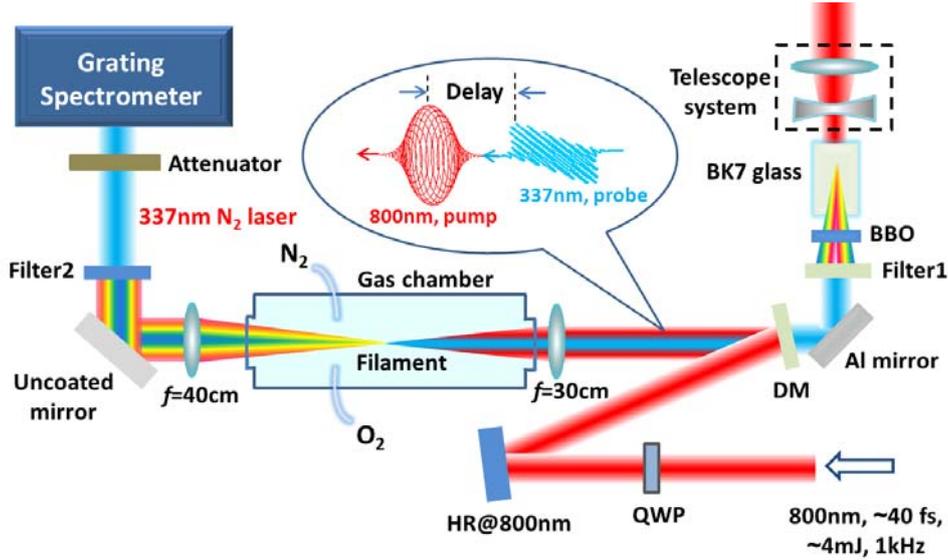

Fig. 1. Schematic diagram of the experimental setup.

## 3. Results and discussion

*3.1 Manifestation of the seeding effect*

Figure 2(a) shows a typical forward 337-nm laser spectrum (red solid line) generated by focusing both the pump and seed pulses in nitrogen gas at one atmospheric pressure (1 atm). The 337-nm laser line corresponds to the transition between the two triplet states $C^3\Pi_u$ ($v'$=0) and $B^3\Pi_g$ ($v$=0) of $N_2$. In this measurement, the delay time between the pump and probe pulses was optimized for obtaining the strongest laser signal. The pump pulse was nearly circularly polarized. For comparison, the spectra of the pump (green dot line) and the probe pulses (blue dash line) are also shown in Fig. 2(a). It can be clearly seen from the three spectra in Fig. 2(a) that the strong laser signal at 337 nm disappears when either the 800-nm pump beam or the 337-nm probe beam is blocked under our experimental conditions. The polarization of the strong 337-nm nitrogen laser pumped by circularly polarized laser pulses was measured by placing a Glan-Taylor polarizer in front of the spectrometer. As shown in the Fig. 2(b), the nitrogen laser at 337 nm has a nearly perfect linear polarization with the polarization direction parallel to that of the probe pulse, which is different from the random polarization of the ASE-type laser [24]. These observations indicate that the probe pulse serves as a seed source to activate the strong stimulated emission. To provide further evidence on the seeding effect, we have also investigated the variation of intensity of the nitrogen laser at 337 nm with the varying central wavelength of the probe pulses, as shown in Fig. 3. It can be clearly seen that when the central wavelength of the probe pulses is 336.9 nm (i.e., close to 337 nm), the nitrogen laser signal becomes the strongest. When the central wavelength of the probe pulses is tuned away from 337 nm, the nitrogen laser signal is significantly reduced. The strong dependence of the 337-nm nitrogen laser signal on the wavelength of the probe pulses provides another evidence for the seed amplification mechanism of the 337-nm laser generation in our experiments.

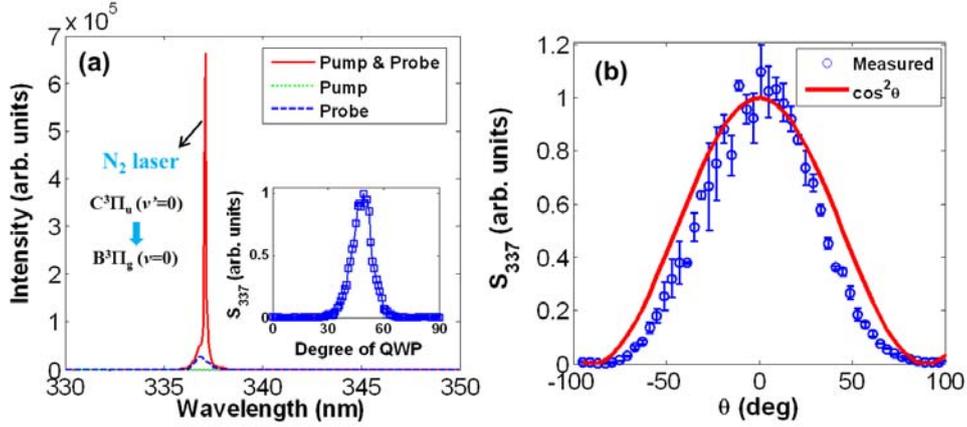

Fig. 2. (a) A typical forward 337-nm laser spectrum generated by focusing both the pump and probe pulses in 1-atm nitrogen gas (red solid line). For comparison, the spectra of the pump and probe pulses are indicated by green dotted line and blue dashed line, respectively. Inset: The change of the nitrogen laser intensity with the polarization of pump pulses. (b) Polarization property of the 337-nm nitrogen laser which is examined by placing a Glan-Taylor prism before the spectrometer. Here $\theta$ is the angle between the optical axis of Glan-Taylor prism and the polarization direction of the 337-nm nitrogen laser.

The seeded 337-nm nitrogen laser shows the characteristic of high sensitivity to the polarization of the pump pulses. The polarization of the pump pulses can be continuously changed from linear to circular using the QWP. As shown in the inset of Fig. 2(a), 337-nm signal can be effectively generated only in a small angle range close to circular polarization. Here, the angles of 0° and 90° correspond to linear polarization, while the angle of 45° corresponds to circular polarization. The other angles correspond to elliptical polarization. As discussed in Ref. [21], this feature suggests that the population inversion between the $C^3\Pi_u$ ($v'=0$) and $B^3\Pi_g$ ($v=0$) states should be achieved by electron-impact excitation due to the higher average kinetic energy of electrons produced in a circularly polarized laser field than in a linearly polarized laser field. However, it is also worth mentioning that a model for quantitatively describing the pumping efficiency of $N_2$ as functions of pump laser intensity and polarization is still lacking, which deserves further investigations.

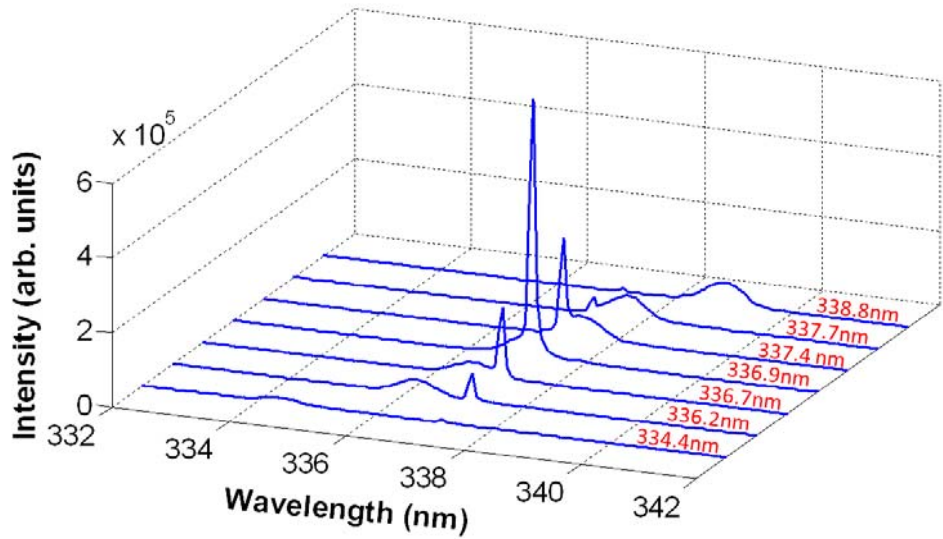

Fig. 3. The signal intensity of 337-nm nitrogen laser as a function of the central wavelengths of the probe pulses, which are indicated on the right side of the corresponding spectra.

*3.2 Gain dynamics of 337-nm laser at various nitrogen gas pressures*

Since the externally seeded nitrogen laser produces much stronger signals than that of the ASE-type laser, the 337-nm nitrogen laser can be generated even at very low gas pressures as shown in Fig. 4(a). This facilitates a systematic investigation on the dependence of nitrogen laser signal on the gas pressure, at which the laser signal appears. It is worth mentioning that at a gas pressure below ~100 mbar, the seeded nitrogen laser disappears. As demonstrated by the curve in Fig. 4(a), in the range of gas pressure from ~100 mbar to ~600 mbar, the laser signal at ~337 nm shows a rapid growth with the increasing gas pressure, followed by the saturation region from ~600 mbar, and then a slow decrease above 900 mbar. The pressure-dependent result can be explained qualitatively based on two processes as below. On the one hand, with the increase of the gas pressure, the densities of both nitrogen molecules and electrons increase, resulting in a higher probability for the inelastic collision between the electrons and nitrogen molecules (i.e., collisional excitation of molecules). On the other hand, at the high gas pressures, collisional quenching between molecules becomes efficient, leading to the reduction of lasing efficiency. The saturation region may reflect the balance between the above two processes.

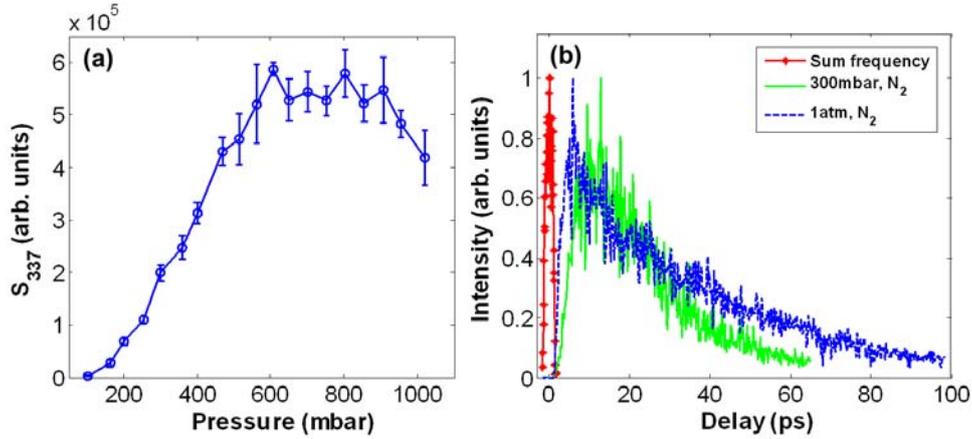

Fig. 4. (a) The intensity of 337-nm nitrogen laser as a function of gas pressures. (b) The temporal evolution of the 337-nm nitrogen laser with the time delay between the pump and probe pulses in the 300-mbar (green solid line) and 1-atm $N_2$ (blue dashed line).

To retrieve the temporal information during the establishment of the population inversion of $N_2$ enabled by the circularly polarized femtosecond laser pulses, we measured the 337-nm nitrogen laser signal as a function of the time delay between the pump and probe pulses, as shown in Fig. 4(b). In this measurement, the zero delay between the pump and probe pulses was set to the instant at which the sum frequency signal of the ~800-nm pump pulse and the ~337-nm probe pulse in a 4-mm-thickness BBO crystal reached the maximum. As indicated by the red asterisk line in Fig. 4(b), the sum frequency signal at ~240 nm has a pulse width of about 2 ps, which corresponds to the pulse duration of the probe pulse because of the relatively short pulse duration of the pump pulse (i.e., ~40 fs). The positive delay in Fig. 4(b) means that the probe pulse is after the pump pulse. The blue dash and green solid lines in Fig. 4(b) show the intensities of the 337-nm laser signals as a function of the pump-probe delay at 1-atm and 300-mbar nitrogen gas, respectively. It can be seen that in both cases, the laser signals appear after the pump pulse, but the 337-nm laser signal increases much faster at 1 atm than at 300 mbar. The rise times of the 337-nm laser signal (defined as the time from the zero delay to the delay time when the maximum lasing signal is measured) are ~6 ps and ~10 ps for the gas pressures of 1 atm and 300 mbar, respectively. The result suggests that the population inversion can be created on a shorter timescale at a higher gas pressure probably due to a higher electron-molecule collision rate. It can also be noticed from Fig. 4(b) that after a rapid growth, the 337-nm laser signal exhibits a slow decay for both the gas pressures, indicating the relaxation process of the population inversion. The 337-nm signal decays slower with the increasing pump-probe time delay at 1 atm than at 300 mbar, which is different from the general behaviour of the fluorescence decay, for which the decay is generally faster at a higher pressure [25]. This may indicate that for free-space nitrogen lasers, both the upper and lower states in the population inversion are strongly dependent on the gas pressures. As a consequence, the generation of the 337-nm nitrogen laser critically depends on the environment in the interaction region, which determines the collisional rate.

*3.3 Influence of molecular oxygen on the 337-nm nitrogen laser*

For remote sensing applications, the free-space 337-nm nitrogen laser eventually needs to be generated in ambient air rather than in pure nitrogen gas. Air contains 21% oxygen gas ($O_2$). Therefore, in order to examine how molecular oxygen affects the generation of the 337-nm nitrogen laser, we measured the signal intensity of the 337-nm nitrogen laser as a function of the concentration of $O_2$ in the mixture gas with the pressure of nitrogen gas being fixed at 800 mbar, as shown in Fig. 5(a). It can be clearly seen that the 337-nm laser signal decreases significantly as the concentration of $O_2$ increases in the gas mixture. Such a strong dependence on the concentration of $O_2$ indicates that oxygen molecules generate a strong quenching effect on the 337-nm nitrogen laser. For comparison, in the inset of Fig. 5(a), we present the laser spectra at 337 nm obtained in pure $N_2$ and in air at the pressure of 1 atm, while all the other experimental conditions are same. Apparently, the intensity of the nitrogen laser in air is about one order of magnitude lower than that in pure nitrogen due to the quenching effect of molecular oxygen.

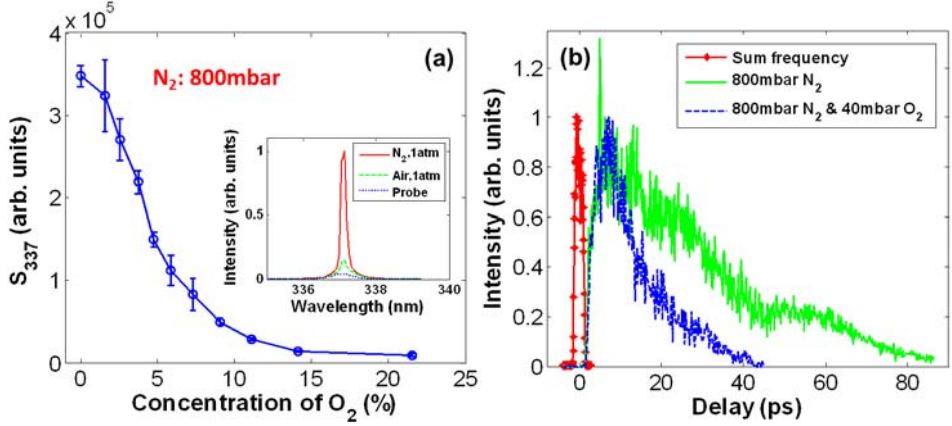

Fig. 5. (a) The signal intensity of 337-nm nitrogen laser as a function of the concentration of $O_2$ in the mixture gas with the pressure of nitrogen gas being fixed at 800 mbar. Inset: The forward laser spectra measured in 1-atm $N_2$ (red solid line) and 1-atm air (green dashed line). For comparison, the spectrum of the probe pulse in air is indicated by blue dotted line. (b) The temporal evolution of the 337-nm nitrogen laser with the varying time delay between the pump and probe pulses in the 800-mbar $N_2$ (green solid line) and in the mixture gas of 800-mbar $N_2$ and 40-mbar $O_2$ (blue dashed line).

To gain a deeper insight on the dynamical process of the collision quenching, Fig. 5(b) compares the temporal evolution of the population inversion in 800 mbar $N_2$ with that in the mixture gases of 800-mbar $N_2$ and 40-mbar $O_2$. Here, the definition of zero delay is the same as that in Fig. 4(b). It can be seen that the addition of 40-mbar $O_2$ leads to a much faster decay (blue dash line) of the population inversion than that in pure nitrogen (green solid line). Generally speaking, influence of molecular oxygen on the generation of the nitrogen laser may be two-fold. On the one hand, since molecular oxygen has a lower ionization potential than that of molecular nitrogen, it should be more easily ionized in strong laser field particularly in tunnel ionization regime. This will lead to a decrease in the laser peak intensity clamped in the filament when compared with the situation of focusing the femtosecond laser in pure nitrogen [26,27]. In this case, the cross section of collision excitation between the electrons and nitrogen molecules will be reduced due to the decrease of kinetic energy of the electrons [28]. On the other hand, the collision reaction between oxygen molecules and nitrogen molecules leads to the de-excitation of nitrogen molecules at the excited state $C^3\Pi_u$ into the ground state [25,29], which accelerates the decay of the population inversion. It has been reported in Ref. [21] that the optical pumping process in nitrogen is extremely sensitive to the presence of molecular oxygen because of a high quenching rate of population inversion in nitrogen-oxygen collision [29], which qualitatively agrees with our observation. Therefore, collisional reaction between oxygen molecules and nitrogen molecules at the excited state $C^3\Pi_u$ is more likely to be the dominant mechanism responsible for a faster decay of nitrogen lasers in the mixture gas.

**4. Conclusion**

In conclusion, we have observed externally seeded laser emissions at 337 nm from neutral nitrogen molecules in a femtosecond laser filament pumped by an intense circularly-polarized pulse. The population inversion created between the $C^3\Pi_u$ and $B^3\Pi_g$ states in nitrogen molecules is ascribed to the collision excitation by the high-energy electrons. The free-space nitrogen laser shows a strong dependence on the wavelength of probe pulses as well as the polarization of pump pulses. In particular, with a pump-probe scheme, the dynamical process of the nitrogen laser generation is revealed. We found that the population inversion reaches its maximum in less than 10 ps, followed by a slow decay in a few tens of picoseconds. Moreover, it is also found that oxygen molecules strongly influence the gain dynamics of the 337-nm laser and significantly decrease its signal intensity due to the collisional quenching effect. Our investigation provides dynamical information for optimizing the generation of the free-space nitrogen laser, which will be useful for generating strong air lasers for remote sensing applications.

**Acknowledgments**

We gratefully acknowledge fruitful discussion with Y. Liu of ENSTA ParisTech/CNRS/Ecole Polytechnique. This work is supported by the National Basic Research Program of China (Grant No. 2011CB808100, No. 2014CB921300), National Natural Science Foundation of China (Grant Nos. 11127901, 11134010, 61221064, 61235003, 11204332 and 11304330), Open Fund of the State Key Laboratory of High Field Laser Physics (SIOM), and Research Fund for the Doctoral Program of Higher Education of China.